\pdfoutput=1

\documentclass[aip,rsi,amsmath,amsymb,reprint, letterpaper]{revtex4-1}

\usepackage{graphicx}
\usepackage{dcolumn}
\usepackage{bm}
\usepackage{hyperref}
\hypersetup{
  colorlinks   = true, 
  urlcolor     = blue, 
  linkcolor    = blue, 
  citecolor   = blue 
}

\usepackage{url} 
\usepackage{upgreek}
\usepackage{color}
\newcommand{\physrep}{Physics Reports}
\newcommand{\Pra}{Physical Review A}
\newcommand{\Prb}{Physical Review B}

\definecolor{navyblue}{rgb}{0.0, 0.0, 0.5}
\definecolor{deepmagenta}{rgb}{0.8, 0.0, 0.8}
\definecolor{dolla-bill}{rgb}{0.52, 0.73, 0.4}

\begin{document}

\title{Broadband loop gap resonator for nitrogen vacancy centers in diamond}

\author{E. Eisenach}
\email{eisenach@mit.edu}
 \affiliation{Department of Electrical Engineering and Computer Science, Massachusetts Institute of Technology, Cambridge, MA 02139, USA}
  \affiliation{MIT Lincoln Laboratory, Lexington, MA 20421, USA}

\author{J. Barry}
\author{R. Rojas}
\author{L. Pham}
 \affiliation{MIT Lincoln Laboratory, Lexington, MA 20421, USA}
\author{D. Englund}
 \affiliation{Department of Electrical Engineering and Computer Science, Massachusetts Institute of Technology, Cambridge, MA 02139, USA}

\author{D. Braje}
 \affiliation{MIT Lincoln Laboratory, Lexington, MA 20421, USA}
\date{\today}

\begin{abstract}
We present an S-band tunable loop gap resonator (LGR) providing strong, homogeneous, and directionally uniform broadband microwave (MW) drive for nitrogen-vacancy (NV) ensembles.  With 42 dBm of input power, the composite device provides drive field amplitudes approaching 5 G over a circular area $\gtrsim\!50$ mm$^2$ or cylindrical volume $\gtrsim\!250$ mm$^3$. The wide 80 MHz device bandwidth allows driving all eight NV Zeeman resonances for bias magnetic fields below 20 G. For pulsed applications the device realizes percent-scale microwave drive inhomogeneity;  we measure a fractional root-mean-square inhomogeneity $\sigma_\text{rms}\!=\! 1.6\%$ and a peak-to-peak variation $\sigma_\text{pp}\!=\! 3\%$ over a circular area of 11 mm$^2$, and $\sigma_\text{rms} \!=\! 3.2\%$ and $\sigma_\text{pp}\! =\! 10.5\%$ over a larger 32 mm$^2$ circular area.  We demonstrate incident MW power coupling to the LGR using multiple methodologies: a PCB-fabricated exciter antenna for deployed compact bulk sensors and an inductive coupling coil suitable for microscope-style imaging. The inductive coupling coil allows for approximately $2\pi$ steradian combined optical access above and below the device, ideal for envisioned and existing NV imaging and bulk sensing applications. 

\end{abstract}




\maketitle

\section{Introduction}\label{intro}

The nitrogen-vacancy (NV) defect center in diamond is employed in a number of wide-ranging applications from quantum information processing~\cite{gaebel2006room,dutt2007quantum} to tests of fundamental physics \cite{waldherr2011violation,hensen2016loophole} to quantum sensing and metrology. In particular, NV-based quantum sensors have demonstrated utility in a broad variety of modalities, including magnetometry~\cite{taylor2008high}, electrometry~\cite{dolde2011electric,chen2017high}, nanoscale NMR~\cite{devience2015nanoscale,kehayias2017solution,bucher2017high}, single proton and single protein detection~\cite{sushkov2013magnetic,lovchinsky2016nuclear}, thermometry~\cite{neumann2013high,kucsko2013nanometre}, time-keeping~\cite{hodges2013timekeeping}, and more~\cite{barry2016optical,laraoui2015imaging,tetienne2017quantum}. Each of these applications takes advantage of one or more principal features of the NV center: all-optical initialization and readout~\cite{vanoort1988optically,nizovtsev2003nv}, long coherence time under ambient conditions~\cite{balasubramanian2008nanoscale,bargill2013solid,myers2017double,bauch2018ultralong}, nanoscale size~\cite{degen2008scanning,maletinsky2012arobust}, or fixed crystallographic axes~\cite{maertz2010vector,clevenson2018robust,schloss2018simultaneous}. However, with notably few exceptions~\cite{wickenbrock2016microwave,akhmedzhanov2017microwave}, \textit{all} NV applications rely on the ability to coherently manipulate the NV ground-state spin via resonant microwave (MW) driving. A number of these applications additionally require generation of strong and uniform MW fields over large areas ($\gtrsim 10$ mm$^2$) or volumes ($\gtrsim 30$ mm$^3$)~\cite{lesage2013optical, glenn2017micrometer, fu2014solar, wolf2015subpicotesla, clevenson2015broadband}, a difficult task that benefits significantly from improvements to standard MW delivery methods. In this work, we discuss the design considerations for a suitable MW delivery mechanism, fabricate a hole-and-slot type loop gap resonator (LGR), and evaluate its performance for NV applications.

\begin{figure}[b!]
\centering
\includegraphics[scale = 0.155]{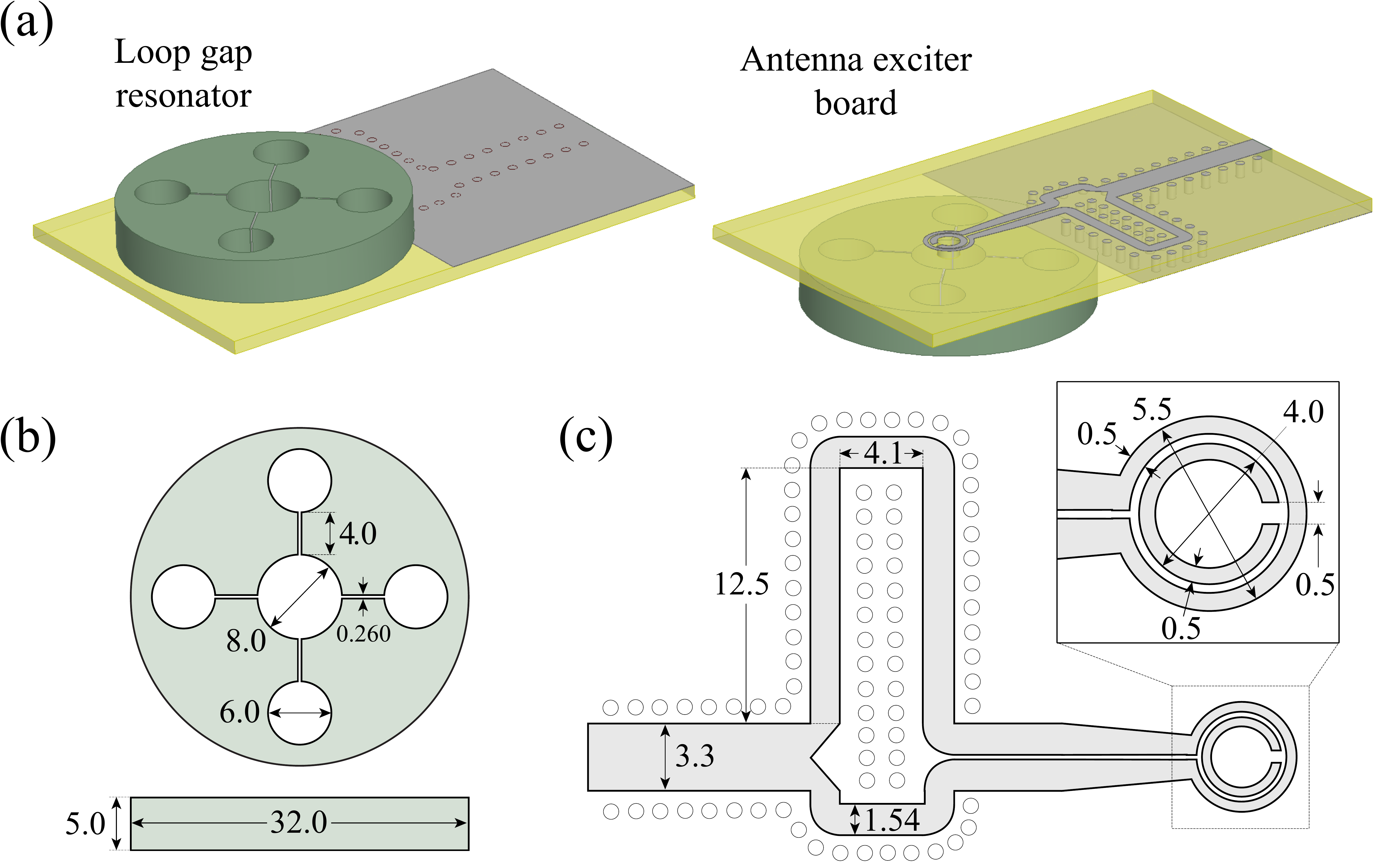}  
\caption{(Color online) \textbf{Loop gap resonator and exciter antenna board.} \textbf{(a)} The metallic resonator employs a five-loop four-gap architecture. Microwaves are coupled into the LGR via the exciter antenna, which is fabricated on a printed circuit board. \textbf{(b)} Line drawing of LGR.   \textbf{(c)} Exciter antenna. A feedline, 50:50 power splitter, and balun (\textbf{bal}anced \textbf{un}balanced) feed the split ring resonator, which is coupled to the LGR. All dimensions are in mm. Optional mounting holes and radial access port for laser excitation are not shown.}
\label{Fig_one}
\end{figure}

Multi-channel imagers and highly sensitive, single-channel bulk sensors are two examples of application modalities that benefit significantly from large detection areas and volumes, respectively. In the case of multi-channel imagers, increasing the detection area extends the measurement field-of-view, whereas for bulk sensors, increasing the detection volume can considerably enhance measurement sensitivity. For example, the shot-noise-limited sensitivity of an NV magnetometer is approximately given by~\cite{taylor2008high}
\begin{equation} \label{equ_1}
\eta \approx \frac{\hbar}{\text{g}_{\text{s}} \upmu_\text{B}} \frac{1}{C\sqrt{\beta \tau}} \frac{1}{\sqrt{N}},
\end{equation}
where $N$ is the number of NV sensors, $\tau$ is the duration of the measurement, $C$ is the measurement contrast, $\beta$ is the number of photons collected per NV per measurement, $\upmu_\text{B}$ is the Bohr magneton, $\text{g}_\text{s} \approx 2$ is the ground state NV\textsuperscript{-} Land\'e g-factor, and $\hbar$ is the reduced Planck constant. The magnetic sensitivity can be improved by increasing $N$, achievable through higher NV density or larger detection volumes. However, NV ensemble coherence times, which limit the optimal measurement time $\tau$, depend inversely on NV density~\cite{taylor2008high}. As a result, there is a practical upper bound on the NV density after which further sensitivity enhancements are attained by increasing the detection volume. For application modalities such as those discussed above, the MW field requires both high power and uniformity in order to achieve high-fidelity quantum state manipulation over the full measurement region.

Standard approaches to applying MW drive to NV ensembles or other solid state spin systems include shorted coaxial loops \cite{clevenson2015broadband,chipaux2015magnetic}, microstrip waveguides \cite{andrich2017long,zhang2016microwave}, coplanar waveguides~\cite{zhang2018vector}, and other coaxial transmission line approaches \cite{mrozek2015circularly}. While such broadband approaches allow abitrary drive frequency, the lack of resonant enhancement forces a compromise between the volume addressed (assuming a fixed homogeneity is required) and MW field strength, denoted $B_1$. Planar lumped-element resonators such as split-ring resonators  \cite{bayat2014efficient,zhang2016microwave}, planar-ring resonators \cite{sasaki2016broadband,twig2010sensitive}, omega resonators~\cite{twig2010sensitive,horowitz2012electron,horsley2018microwave,simpson2017electron}, and patch antennas \cite{zhang2016microwave} forego the flexibility of broadband solutions in favor of resonantly enhanced magnetic fields, thus enabling MW driving over larger regions. For example, the split-ring resonator presented by Bayat $\textit{et al.}$ achieves a MW field strength of $B_1= 5.6$ G and a fractional root-mean-square inhomogeneity of $\sigma_\text{rms}\approx 4.4\%$ over a $\sim\!1$ mm$^2$ area \cite{bayat2014efficient}. However, such planar structures are ill-suited to providing good $B_1$ homogeneity away from the plane of fabrication. The community has addressed this shortcoming by employing a variety of three-dimensional resonators. Enclosed metallic cavity resonators \cite{rose2017coherent}, enclosed dielectric resonators \cite{breeze2017continuous,floch2016towards,creedon2015strong}, open dielectric resonators \cite{kapitanova2017dielectric}, and certain three-dimensional lumped element resonators \cite{angerer2016collective} all allow for good homogeneity over large volumes but unfortunately offer little to no optical access. As all-optical initialization and readout is a primary benefit for many solid-state spin systems, including NV-diamond \cite{doherty2013nitrogen}, such a trade-off is incompatible with many existing and envisioned applications~\cite{schirhagl2014nitrogen}. 


To address this current shortcoming we present a three-dimensional tunable loop gap resonator. The design is based on the anode block of a hole-and-slot-type cavity magnetron and, similar to certain devices discussed above, utilizes resonant enhancement to achieve the desired MW drive strengths over large areas ($>$50 mm$^2$) or volumes ($>$250 mm$^3$). The design has an open geometry; for interrogation volumes centered within the LGR, approximately half of the $4\pi$ solid angle remains optically accessible. Importantly, for currently semi-standardized commercial diamond plates (2-4.5 mm side lengths with 0.5 mm thickness) this solution allows maximal access to the diamond's large front and back faces. The open access, good homogeneity, and high $B_1$ fields over the 8 mm diameter by 5 mm thickness cylindrical volume make the device well-suited both for wide-field magnetic imaging---applicable to studies of living systems \cite{kucsko2013nanometre,lesage2013optical,barry2016optical,davis2018mapping}, early earth rocks or meteorites \cite{glenn2017micrometer,fu2014solar}, single cells~\cite{glenn2015single}, electronic devices~\cite{simpson2016magneto}, etc.---and for single-channel bulk sensing \cite{acosta2009diamonds,wolf2015subpicotesla,clevenson2015broadband,chatzidrosos2017miniature,barry2016optical} targeting geosurveying, magnetic anomaly detection, space weather monitoring, etc.


\section{Loop Gap Resonator Design and Fabrication}

\begin{figure}[t!]
\centering
\includegraphics[scale = 0.25]{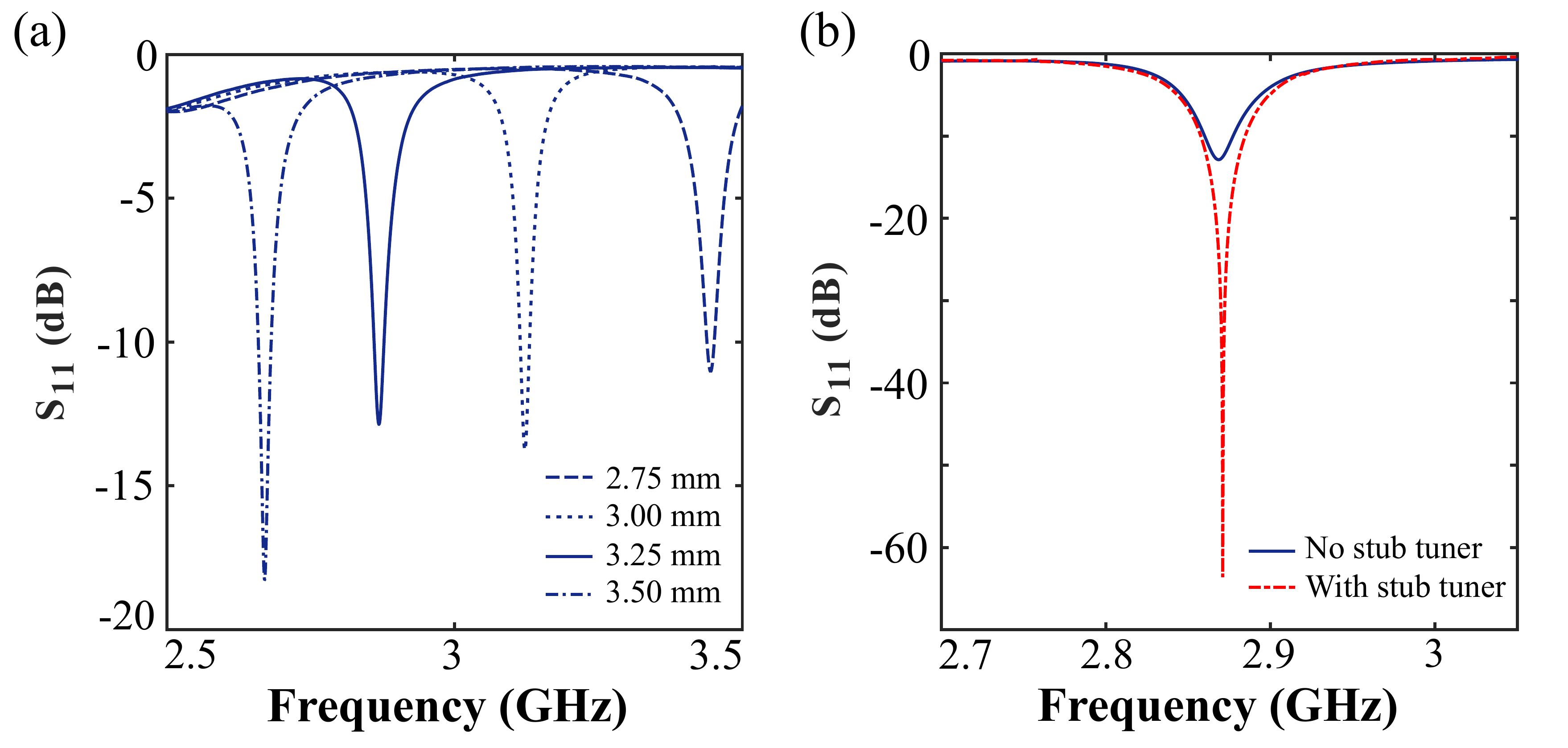}  
\caption{(Color online) \textbf{Frequency tuning and impedance matching of LGR composite device.}  \textbf{(a)} The resonant frequency $f_\text{0}$ is adjusted by translating the sapphire shims in the four capacitive gaps. In the absence of a stub tuner, the LGR composite device exhibits $S_{11}$ values between -10 and -20 dB from 2.5 to 3.5 GHz, indicating $\gtrsim\!90\%$ of power delivered to the LGR composite device contributes to $B_1$ in this range. \textbf{(b)} Nearly perfect critical coupling can be achieved with a stub tuner, allowing practically all incident MW power to contribute to $B_1$.}  \label{Fig_two}
\end{figure}

A standard hole-and-slot LGR with $n$ outer loops may be approximated as $n$ coupled LC resonators oscillating in tandem~\cite{wood1984loop}. Circulating currents around the central and outer loops create a total inductance $L$, given by \cite{froncisz1982loop,froncisz1986q,wood1984loop}
\begin{equation} \label{equ_2}
L \approx \frac{L_c n L_o}{n L_o+L_c},
\end{equation} 
where $L_c$ and $L_o$ denote the inductance of the central loop and of a single outer loop, respectively. Similarly, the $n$ narrow capacitive gaps create a total capacitance C, which is given by \cite{froncisz1982loop,froncisz1986q,wood1984loop}
\begin{equation} \label{eqn_3}
C \approx \frac{\epsilon_r \epsilon_0 A}{n d},
\end{equation}
where $A$ and $d$ are the capacitive gap side wall area and separation, respectively. The resonant frequency of the LGR is therefore given by
\begin{equation} \label{eqn_4}
f_{0} = \frac{1}{2\pi \sqrt{L C}}.
\end{equation}

In practice, the central loop diameter is set to $\sim$5-10 mm, corresponding to the typical size of a diamond plate, whereas $d$ is limited by practical machining tolerances and $\epsilon_r$ by physically available materials. The capacitive gap area $A$ is constrained by the dual LGR design objectives of (i) maintaining optical accessibility, which limits the thickness of the LGR device, and (ii) bounding $f_0$ above the target resonant frequency in order to allow for further tuning via shims (discussed below). Additionally, while increasing the number $n$ of loops and gaps can improve $B_1$ uniformity~\cite{piasecki1993field}, this approach results in a denser mode spectrum \cite{froncisz1982loop} and increases the likelihood of cross-mode excitations deleteriously altering the field distribution within the central loop. As a compromise, our design employs $n=4$ outer loops [Fig. 1(b)], thus allowing for sufficient uniformity 
while locating the closest eigenmode more than 1.5 GHz below the TE$_{01}$ eigenmode. 

The LGR detailed in this work consists of a central loop of radius $r_c = 4$ mm surrounded by $n=4$ symmetrically arranged outer loops of radius $r_o=3$ mm, as shown in Fig. 1(b). The outer loops return magnetic flux to the central loop and therefore oscillate antisymmetrically with the central loop (180$^\circ$ out of phase). The side walls of the capacitive gaps are separated by $d = 260$ $\upmu$m.  With these dimensions, Eqns.~\ref{equ_2} and \ref{eqn_3} predict $L = 8.7$ nH and $C = 0.17$ pF respectively, resulting in an expected resonant frequency for the naked air-gapped LGR of $f_0 = 4.1$ GHz, approximately 1.2 GHz above the NV resonance frequencies. For comparison, the measured $f_0$ for the air-gapped resonator is in the 4.6-4.9 GHz range. 

The LGR resonant frequency $f_0$ is additionally tuned by inserting and translating dielectric shims in the LGR's capacitive gaps, thereby increasing total capacitance $C$ until $f_0$ overlaps the NV resonance frequencies as desired. As shimming material, we employ 200 $\upmu$m thick C-plane sapphire, which is commercially available in semiconductor grade 50.8 mm diameter wafers, can be cut on standard wafer dicing saws, has a high relative permittivity of $\epsilon_r =11.5$ parallel to the C-plane  \cite{westphal1972dielectric} (allowing for a large tuning of $f_0$), and exhibits low dielectric loss ($\text{Tan}\,\delta\!< .0001$ at 3 GHz \cite{westphal1972dielectric,hartnett2006sapphireshim}). The sapphire shims are cut to lengths longer than the $l_c=4$ mm radial length of the capacitive gaps and wedged into the $n=4$ capacitive gaps with teflon thread tape. These sapphire shims are then translated radially until the desired value of $f_0$ is attained. The shims are always positioned so that excess shim length extends into the outer rather than the central loop, in order to minimally perturb the central loop $B_1$ field.  Simulations further suggest that radially symmetric shim configurations produce the best $B_1$ field homegoneity, as asymmetries in shim placement perturb the desired TE$_{01}$ field distribution. Insertion and removal of diamonds in the LGR composite device typically leaves $f_0$ unchanged, as the large electric fields of the TE$_{01}$ mode are predominantly confined to the capacitive gaps (see Appendix \ref{electricfields}). 


The LGR is fabricated via wire electron discharge machining, which is well-suited for producing the tight tolerances and vertical side walls required for the narrow $d=260$ $\upmu$m capacitive gaps. A  titanium alloy (Ti-6Al-4V) was chosen as the resonator cavity material. The lower conductivity of this alloy compared to that of copper ($\sigma_\text{Ti} = 5.7\pm  0.1 \times 10^5$ S/m vs. $\sigma_\text{Cu} =5.9 \times 10^7$ S/m) allows for a broader resonance with a 3dB bandwidth $\Delta_\text{3dB}=80$ MHz, sufficient to address all eight NV resonances for bias magnetic fields $B_0$ up to $\sim\!20$ G. This 80 MHz bandwidth corresponds to a loaded quality factor $Q_L \equiv f_0/\Delta_\text{3dB} \approx 36$ when the LGR is critically coupled to the driving source. The LGR may be optionally fit with a radial  access hole (for laser excitation of the NV ensemble) and three $\#$2-56 mounting holes, which affix the LGR to an exciter antenna, discussed next. 

\begin{figure}[t!]
\centering
\includegraphics[scale = 0.26]{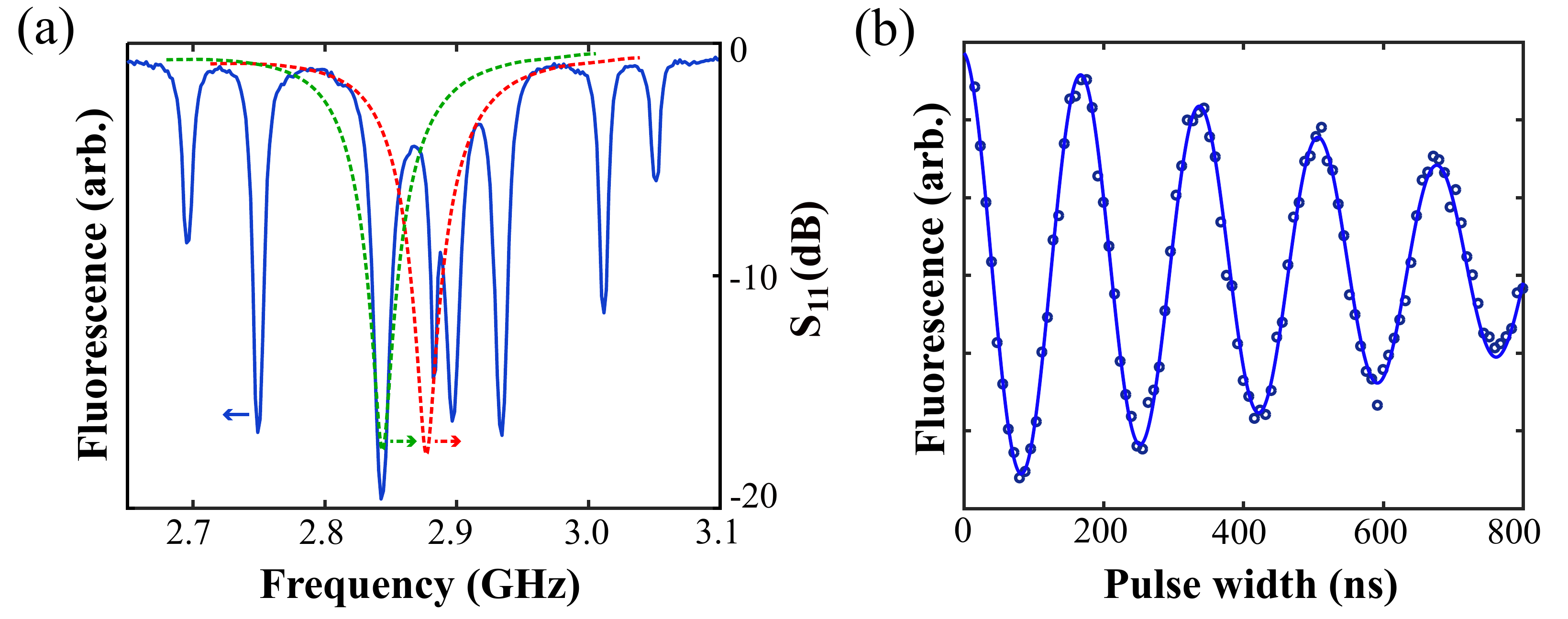}  
\caption{(Color online) \textbf{LGR driving of an NV ensemble}. \textbf{(a)} NV electron spin resonance spectrum (\textcolor{blue}{\textbf{---}}) under application of bias field $B_0$. The bias field allows individual addressing of all eight NV resonances, arising from the combination of the two allowed magnetic dipole transitions with the four possible NV orientations. The NV hyperfine structure is obscured by MW power broadening and the contrast variation between the NV resonances is attributed primarily to the $S_{11}$ line-shape. 
The $S_{11}$ parameter is shown before (\textcolor{red}{\textbf{-\,-\,-}}) and after (\textcolor{dolla-bill}{\textbf{-\,-\,-}}) shifting the LGR resonant frequency $f_0$ to the target NV resonance. Arrows indicate corresponding y axes. \textbf{(b)} Typical data depicting Rabi oscillations under MW excitation at the target NV resonance frequency indicated in (a). Data (\textcolor{navyblue}{$\mathbf{\circ}$}) is fit (\textcolor{blue}{\textbf{---}}) to an exponentially decaying sinusoid. 
}  \label{Fig_three}
\end{figure}

\section{Loop Gap Resonator Coupling and Exciter Antenna Board} \label{exciterantenna}

Incident MW power $P$ is inductively coupled into the LGR by an exciter antenna, composed of a split ring resonator that is differentially driven by a microstrip balun, as shown in Fig. \ref{Fig_one}(c). The differential driving mitigates common-mode noise on the two traces, which might otherwise couple to the split-ring resonator. Although the microstrip balun is designed to match the feed-line and the split ring component of the exciter antenna at frequencies near 2.87 GHz, good matching is achieved from 2.5 GHz to 3.5 GHz as well. For drive frequencies between 2.5 and 3.5 GHz, the exciter antenna board couples more than 90$\%$ of incident MW power into the LGR, as shown in Fig. \ref{Fig_two}(a). For a specific fixed frequency, the impedance matching may be further optimized by inserting a stub tuner between the MW source and the exciter antenna board, as shown in Fig. \ref{Fig_two}(b). 

A via shield along a portion of the balun helps reduce interference and cross-talk between traces, controls the trace impedance, and reduces radiative losses along the balun's $\pi$-phase delay arm. The exciter antenna is fabricated from 1 oz. copper trace with immersion silver finish on 1.524 mm thick dielectric (Rogers, RO4350B). Although the proximity of the split ring resonator perturbs the field distribution inside the LGR, both simulations and measurements suggest this effect is small and not the dominant inhomogeneity source (see Section \ref{FieldHom}). For applications intolerant of such perturbations or those requiring maximal diamond optical access, we achieved similar success inductively coupling a small coil of radius $\approx r_o$ to one of the outer loops \cite{koskinen1992the}, where the coil is translated (via mechanical stage) until the desired coupling is achieved. We expect this coupling method to be particularly suitable for laboratory or clinical imaging applications.


\section{Loop Gap Resonator Performance} \label{FieldHom}

The strength and homogeneity of $B_1$ within the LGR central loop is evaluated employing standard NV techniques, as described in detail in Ref. \cite{pham2013magnetic} and elsewhere \cite{childressthesis2011coherent,mazethesis2010quantum}. A $\{$100$\}$-cut diamond plate containing $\sim 1 \times 10^{14}$ NV/cm$^3$ is mounted at the center of the LGR with the $<$100$>$ crystallographic axis collinear with the LGR axis. A rare earth magnet creates a static magnetic bias field $B_0$, which shifts the energies of the $m_s=\pm1$ ground-state Zeeman sublevels. The energy shifts are given to first order by~\cite{taylor2008high}
\begin{equation}
\Delta E \approx \text{g}_{\text{s}} \upmu_\text{B} m_s \vec{B}_0\cdot \hat{n}_i,
\end{equation}
where $\hat{n}_i$ denotes a unit vector oriented along one of the four diamond crystallographic axes. By judicious choice of $\vec{B}_0$, all eight energy levels and associated $m_s\!=\!0\! \leftrightarrow \!m_s\! =\!\pm1$ magnetic dipole transitions can be isolated as shown in Fig. \ref{Fig_three}(a). The resonator is tuned to excite a single NV transition, yielding Rabi oscillations [Fig. \ref{Fig_three}(b)]. The data is fit to an exponentially decaying sinusoid in order to extract the Rabi frequency $\Omega_R$, from which the magnitude of $B_1$ can be calculated as 
\begin{equation} \label{eqn:rabitob1}
B_1 =\sqrt{3} \frac{\hbar \Omega_R}{\text{g}_{\text{s}} \upmu_\text{B}}.
\end{equation}
\begin{figure}[t!]
\centering
\includegraphics[scale = 0.35]{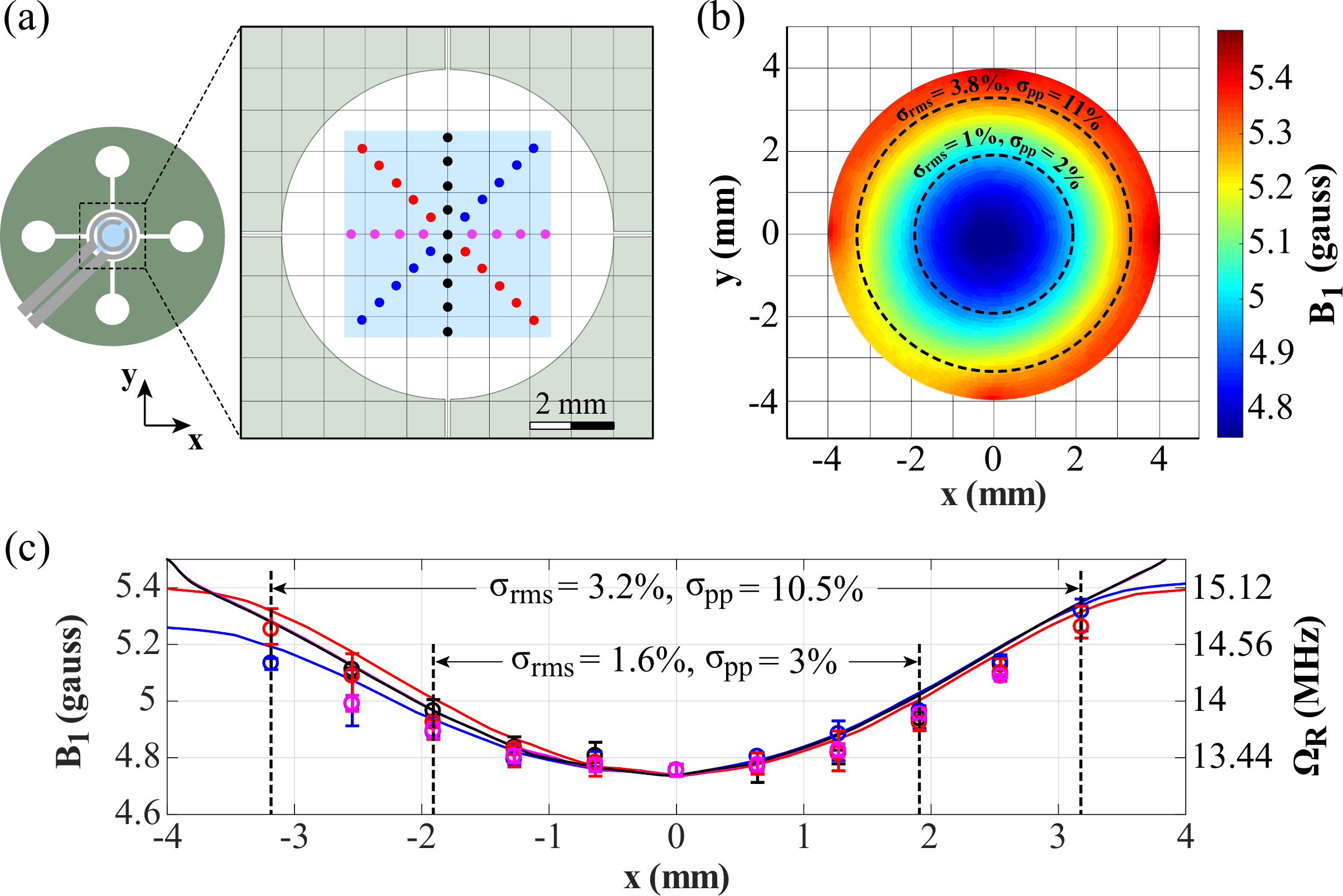}  
\caption{(Color online) \textbf{$\boldsymbol{B_1}$ field uniformity of LGR composite device.} \textbf{(a)} An NV-containing 4.5 mm $\times$ 4.5 mm diamond plate is placed in the LGR central loop, and the  Rabi frequency is measured where indicated (\textcolor{deepmagenta}{\textbullet},\textcolor{black}{\textbullet},\textcolor{red}{\textbullet},\textcolor{blue}{\textbullet}) to characterize $B_1$. \textbf{(b)} Simulations suggest the $B_1$ field distribution should be approximately radially symmetric, with the leading order deviation resulting from the exciter antenna. Dashed lines indicate the 32 mm$^2$ and 11 mm$^2$ areas within which the $B_1$ field uniformity is evaluated. \textbf{(c)} $B_1$ field measurements (\textcolor{deepmagenta}{$\circ$},\textcolor{black}{$\circ$},\textcolor{red}{$\circ$},\textcolor{blue}{$\circ$}) at the points depicted in (a) and simulations (\textcolor{deepmagenta}{\textbf{--}},\textcolor{black}{\textbf{--}},\textcolor{red}{\textbf{--}},\textcolor{blue}{\textbf{--}}) along each locus of points are in good agreement. Error bars indicate 1-sigma uncertainty for the $B_1$ measurement. Dashed lines indicate the radial boundaries of the 32 mm$^2$ and 11 mm$^2$ areas over which $B_1$ field uniformity is evaluated. The measured $B_1$ uniformity is given for each area.}
\label{Fig_four}
\end{figure}
In this geometry, the $B_1$ field is oriented along the [100] crystallographic axis of the diamond, degenerately offset from all four NV axis orientations by half the tetrahedral bond angle $\theta_{\text{tet}}/2 = \text{ArcCos}\frac{1}{\sqrt{3}} \approx 54^\circ$. NV Rabi oscillations are driven by the $B_1$ field component transverse to the NV symmetry axis, reducing the Rabi frequency by $\sqrt{2/3}$ \cite{sasaki2016broadband}. Accounting for the rotating wave approximation introduces another factor of $1/\sqrt{2}$, resulting in the conversion factor $\sqrt{3}$ in Eq. \ref{eqn:rabitob1}. To ensure $\vec{B}_0$ is consistent for all measurements across the LGR central loop [Fig. \ref{Fig_four}(a)], the confocal excitation volume is held fixed with respect to the $B_0$-generating permanent magnet, and the diamond and LGR composite device are translated together. We employ a long working distance objective (Mitutoyo 378-803-3, M Plan Apo 10$\times$ NA=0.28) to collect the NV fluorescence; the 34 mm working distance is necessary to minimize perturbation of the $B_1$ field by the metal objective housing. Future NV wide-field imaging applications may require ceramic-tipped objectives.

Application of incident MW power $P \approx$ 42 dBm yields an axially oriented $B_1$ at the center of the LGR with magnitude 4.7 G. The corresponding Rabi frequency $\Omega_R$ = $2\pi\times 7.7$ MHz for NV centers oriented at half the tetrahedral bond angle relative to the LGR axis. Qualitatively, as shown in Fig. \ref{Fig_four}(c), $B_1$ displays a minimum at the LGR center, increases in magnitude with increasing radial displacement from the center, and is approximately radially symmetric. The best homogeneity is therefore expected at the LGR center.

The $B_1$ field uniformity is quantitatively characterized using both the fractional root-mean-square inhomogeneity $\sigma_\text{rms}$ and the fractional peak-to-peak variation $\sigma_\text{pp} = \big[B_1^\text{max}-B_1^\text{min}\big]/B_1^\text{average}$. The use of both metrics facilitates comparison with alternative existing designs. Over a 32 mm$^2$ circular area axially centered in the LGR central loop, we observe $\sigma_\text{rms} \!=\! 3.2\%$ and $\sigma_\text{pp}\! =\! 10.5\%$, as shown in Fig. \ref{Fig_four}(c). Over a smaller 11 mm$^2$ circular area, we observe $\sigma_\text{rms}\!=\! 1.6\%$ and $\sigma_\text{pp}\!=\! 3\%$.

The LGR performance is modeled using a commercial finite element MW simulation package (Ansys, HFSS). Simulations include the exciter antenna board, which causes a small perturbation to the otherwise radially symmetric field [Fig. \ref{Fig_four}(b)]. The simulation predicts $B_1 \approx$ 4.8 G at the LGR center with incident MW power $P=42$ dBm. Within a 32 mm$^2$ circular area centered in the LGR central loop, simulations indicate $\sigma_\text{rms} \!=\! 3.8\%$ and $\sigma_\text{pp} \!=\! 11\%$, whereas in a smaller 11 mm$^2$ circular area, simulations indicate $\sigma_\text{rms} \!=\! 1\%$ and $\sigma_\text{pp} \!=\! 2\%$. These simulation results are in good agreement with the measurements above. 

As a three-dimensional cavity resonator, the LGR provides better axial field uniformity than planar-only geometries~\cite{floch2016towards,kapitanova2017dielectric,angerer2016collective}. For example, for a 3.14 mm\textsuperscript{3} cylindrical volume (1 mm radius disk with 1 mm thickness), simulations yield $\sigma_{\text{rms}}=0.8\%$, $\sigma_{\text{pp}}=3.7\%$ and an average $B_1$ of 4.8 G (see Appendix \ref{appaxfielduniformity}).

{\section{Discussion}

The device presented here exhibits further benefits which we now discuss, along with extensions tailored for specific applications. For example, for ubiquitously employed pulsed measurement protocols, a short ring-down time $\tau_\text{ring}$ (i.e., $B_1$ field $1/e$ decay time) is necessary for high-fidelity pulse shape control. Although techniques to compensate for long ring-down times are effective~\cite{tabuchi2010total,borneman2012bandwidth,peshkovsky2005rf}, shorter native values of $\tau_\text{ring}$ are nonetheless generally desired~\cite{pfenninger1995general,rinard2005loopgap}. The observed loaded quality factor $Q_L = 36$ corresponds to a ring-down time of $\tau_\text{ring} = 4$ ns (see Appendix \ref{app}), making the device suitable for standard pulsed protocols~\cite{Smeltzer2009Quantum, Jelezko2004Observation}. 

Due to square-root scaling of $B_1$ with incident MW power ($B_1 \propto \sqrt{P}$), higher power handling can allow for stronger $B_1$ fields. The non-planar resonator design allows for otherwise higher incident MW powers as currents circulate over an extended 2D surface (versus the 1D edge for a planar structure). Further, the metallic LGR thermal mass and  thermal conductivity allow efficient heat transfer and sinking, resulting in improved device stability and power handling. Although the latter was not tested, the LGR composite device is expected to allow $>\!$ 100 W for CW and pulsed operation, limited by dielectric breakdown of air in the 260 $\upmu$m capacitive gaps. Should available MW power be constrained, stronger $B_1$ can be achieved by fabricating the LGR from a more electrically conductive material (e.g. silver or copper) at the expense of bandwidth. In such circumstances, the bandwidth can be continuously adjusted above its minimum value by over-coupling the resonator (at the expense of reduced $Q_L$). 

While the presented LGR is 5 mm thick, the fundamental hole-and-slot approach is expected to be feasible for a variety of thicknesses. A thicker device will provide better field uniformity at the expense of optical access. In contrast, for applications requiring MW delivery over a thin planar volume, we expect the LGR can be fabricated via deposition on an appropriate insulating substrate, as discussed in Refs.~\cite{twig2013ultra,twig2010sensitive}. We have found semi-insulating silicon carbide~\cite{schloss2018simultaneous} suitable due to the material's high thermal conductivity ($\approx$490 W/(m*K)~\cite{protik2017phonon,qian2017anisotropic}, high Young's modulus, moderate cost and wide availability in semi-conductor grade wafers. Our simulations suggest the planar LGR approach can offer modest improvements in $B_1$ homogeneity over split ring resonators. 

Although the exciter antenna (see Section \ref{exciterantenna}) facilitates a compact, vibration-resistant, and portable device, this component introduces non-idealities in both field uniformity and optical access. As similar scattering parameters are obtained by inductively coupling a small coil to one of the LGR outer loops, this latter solution may find favor for applications requiring maximal optical access and, furthermore, requires no PCB fabrication.

In this work, we demonstrated a broadband tunable LGR allowing appplication of strong homogeneous MW fields to an NV ensemble. The LGR demonstrates a dramatic improvement over prior MW delivery mechanisms, both improving on and spatially extending MW field homogeneities. We expect the device to be useful for bulk sensing~\cite{acosta2009diamonds,wolf2015subpicotesla,clevenson2015broadband,chatzidrosos2017miniature,barry2016optical} and particularly imaging applications~\cite{karaveli2016modulation,glenn2015single,barry2016optical,lesage2013optical,wu2016diamond,fu2014solar,glenn2017micrometer}, due to the optical access allowed by the LGR composite device both above and below the diamond. 

\section{Acknowledgments}

The authors would like to thank P. Hemmer, M. Newton, C. McNally, and S. Alsid for helpful discussions, and G. Sandy for resonator design simulations. E. Eisenach was supported by the National Science Foundation (NSF) through the NSF Graduate Research Fellowships Program. Any opinions, findings, conclusions, or recommendations expressed in this material are those of the author(s) and do not necessarily reflect the views of the U.S. Government.

\appendix
\section{Appendix A} \label{appA}

\renewcommand{\theequation}{\thesection.\arabic{equation}} 
\setcounter{equation}{0} 

\renewcommand{\thefigure}{\thesection.\arabic{figure}} 
\setcounter{figure}{0} 

{\subsection{The NV Center in Magnetometry}

The negatively-charged NV color center (NV\textsuperscript{-}) is a deep band gap impurity within the diamond crystal lattice [Fig. \ref{ApFig_NV}(a)]. The point defect's $C_{3v}$ symmetry results in a $^3$A$_2$ spin-triplet ground state and a $^3$E spin-triplet excited state, separated by a zero phonon line (ZPL) of 637 nm \cite{maze2011properties}. Spin-spin interactions give rise to a zero-field splitting in the ground-state spin triplet, shifting the $m_s = \pm 1$ states with respect to the $m_s = 0$ state by D\textsubscript{gs} $\approx$ 2.87 GHz [Fig. \ref{ApFig_NV}(b)]. In the presence of a static magnetic field $B_0$, the $m_s = \pm1$ sublevels experience Zeeman splitting proportional to the projection of the magnetic field along the NV symmetry axis. Above-band optical excitation (typically performed with a 532-nm laser) results in phononic relaxation of the NV spin within the $^3$E excited state, followed by fluorescent emission in a broad band. While these optical transitions are generally spin-preserving, an alternate decay path through a pair of metastable singlet states ($^1$A$_1$ and $^1$E) results preferential relaxation from the $m_s = \pm 1$ excited states to the $m_s = 0$ ground state that is non-radiative in the typical $637-800$ nm fluorescence band. This behavior under optical excitation has two major consequences: (1) an optical means of polarizing the NV spin, and (2) optical detection via spin-state-dependent fluorescence intensity.

\begin{figure}[b!]
\begin{center}
	\includegraphics[scale = 0.22]{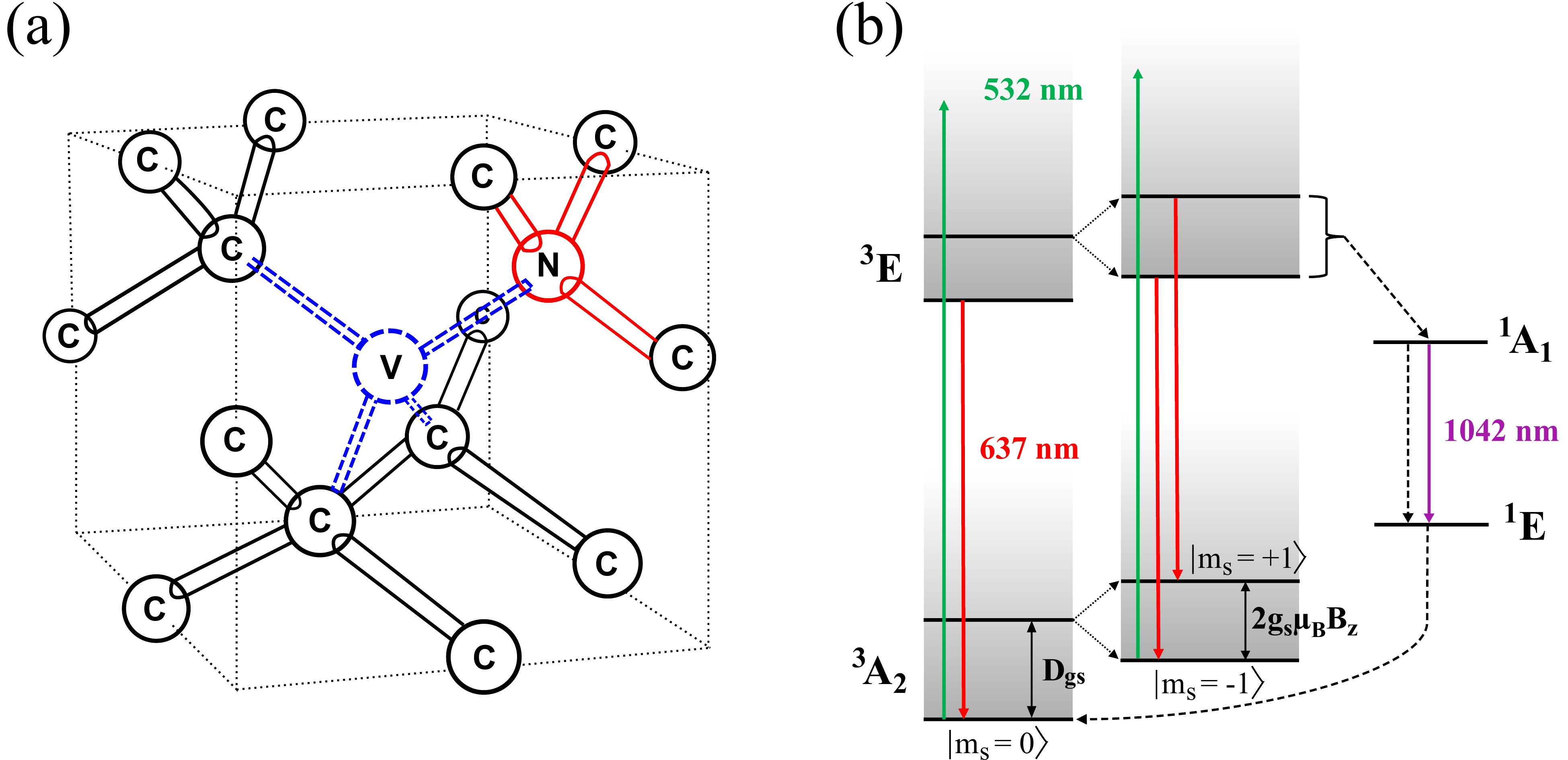}  
\caption{(Color online) \textbf{The NV center structure.} \textbf{(a)} The NV electronic energy level structure. \textbf{(b)} Example of one NV center orientation within the diamond crystal structure.}  \label{ApFig_NV}
\end{center}
\end{figure}

Measurement of the NV electron spin resonance (ESR) spectrum can be performed by sweeping the carrier frequency of the MW drive field and monitoring NV fluorescence in the visible band. Generally, the continuous optical excitation pumps the NV spin population into the more fluorescent $m_s = 0$ state; however, when the carrier frequency is resonant with an NV spin transition, the NV spin population is cycled into an $m_s = \pm 1$ state, causing decreased fluorescence intensity, which appears as a dip in the ESR spectrum \cite{jensen2017magnetometry,rondin2014magnetometry}. Since the NV symmetry axis may be aligned along one of four possible crystal-defined orientations---each orientation being equally thermodynamically likely in low strain diamond---the ESR spectrum can contain up to eight distinct non-degenerate NV resonances, which probe different field components. The different orientations act as basis vectors, which collectively span the space and allow the total vector field to be reconstructed \cite{jensen2017magnetometry}.


\section{Appendix B} \label{app}

\renewcommand{\theequation}{\thesection.\arabic{equation}} 
\setcounter{equation}{0} 

\renewcommand{\thefigure}{\thesection.\arabic{figure}} 
\setcounter{figure}{0} 

\subsection{Cavity Ring-down Time}

The $1/e$ cavity ring down time of the $B_1$ field is \cite{pfenninger1995general,rinard2005loopgap}
\begin{equation}
\tau_\text{ring} = \frac{Q}{\pi f_0}.
\end{equation}
At critical coupling $Q = Q_L = 36$, yielding $\tau_\text{ring}=4$ ns.

\begin{figure}[t!]
\begin{center}
\includegraphics[scale = 0.35]{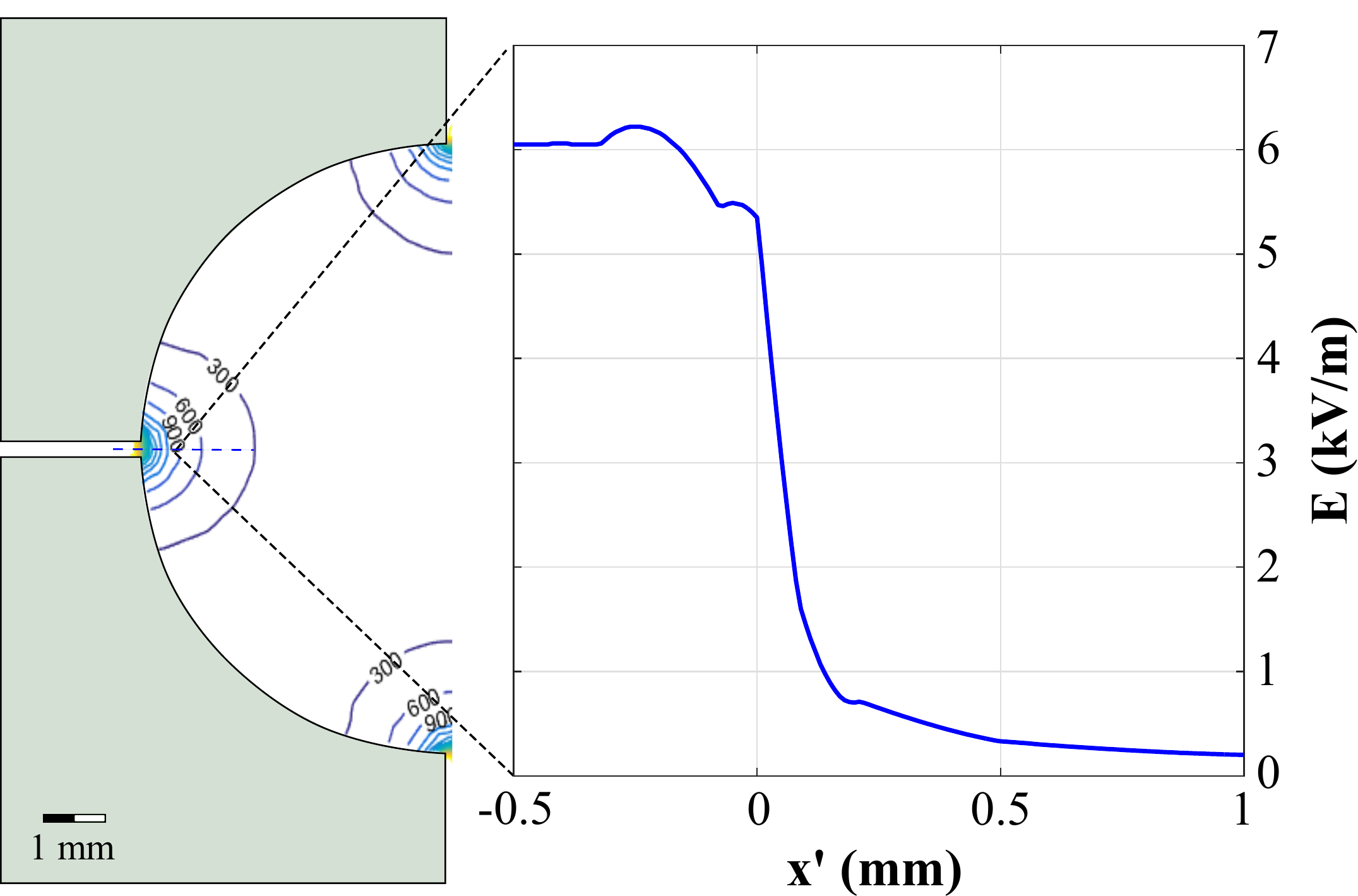}  
\caption{(Color online) \textbf{Simulated electric field magnitude $E$ in vicinity of LGR capacitive gap.} Inset depicts the electic field magnitude $E$ as a function of distance from the capacitive gap with $x'=0$ mm corresponding to the plane of the central loop-gap interface. Observed subfeatures between -0.3 mm and 0.2 mm arise from fringing fields, and are complicated by the 260 $\upmu$m gap only being partially filled by the 200 $\upmu$m thick sapphire.}  \label{ApFig_one}
\end{center}
\end{figure}

\subsection{Electric Field Simulation}\label{electricfields}


Ideally, the electric fields of the TE$_{01}$ cavity mode are completely confined within the LGR capacitive gaps, ensuring that $f_0$ remains constant when differently-sized diamonds are placed within the central loop. In practice, fringing electric fields from the capacitive gaps extend partially into the LGR's central loop as shown in Fig. \ref{ApFig_one}. However, at distances $>$ 1 mm from the capacitive gaps, the electric field magnitude $E$ is decreased by $>\!10\times$ from the peak field inside the capacitive gap. Consequently, insertion of a diamond (with $\epsilon_r \approx 5.7$ at 3 GHz \cite{ibarra1997wide}) beyond this region has little if any effect on the LGR resonant frequency $f_0$.   

\subsection{Axial Field Uniformity}  \label{appaxfielduniformity}

Figure \ref{ApFig_Z} plots the simulated magnitude of $B_1$ along the LGR's symmetry axis, illustrating the improved axial field uniformity possible with three-dimensional cavity resonators~\cite{floch2016towards,kapitanova2017dielectric,angerer2016collective}, compared to that of planar-only geometries. The presence of the split ring resonator at $z = 4.024$ mm perturbs $B_1$ inside the LGR, shifting the point of maximal $B_1$ down by 0.4 mm, away from the split ring resonator. Within a cylindrical volume of 3.14 mm$^3$ (1 mm radius and 1 mm thickness), centered around the point of maximal $B_1$, the simulation predicts $\sigma_\text{rms} = 0.78\%$ and $\sigma_\text{pp} = 3.7\%$. For a larger cylindrical volume of 12.6 mm$^3$ (2 mm radius and 1 mm thickness), the simulation predicts $\sigma_\text{rms} = 2\%$ and $\sigma_\text{pp} = 8\%$. These dimensions are comparable to those of commercially available single-crystal diamonds.

\begin{figure}[h!]
\begin{center}
\includegraphics[scale = 0.43]{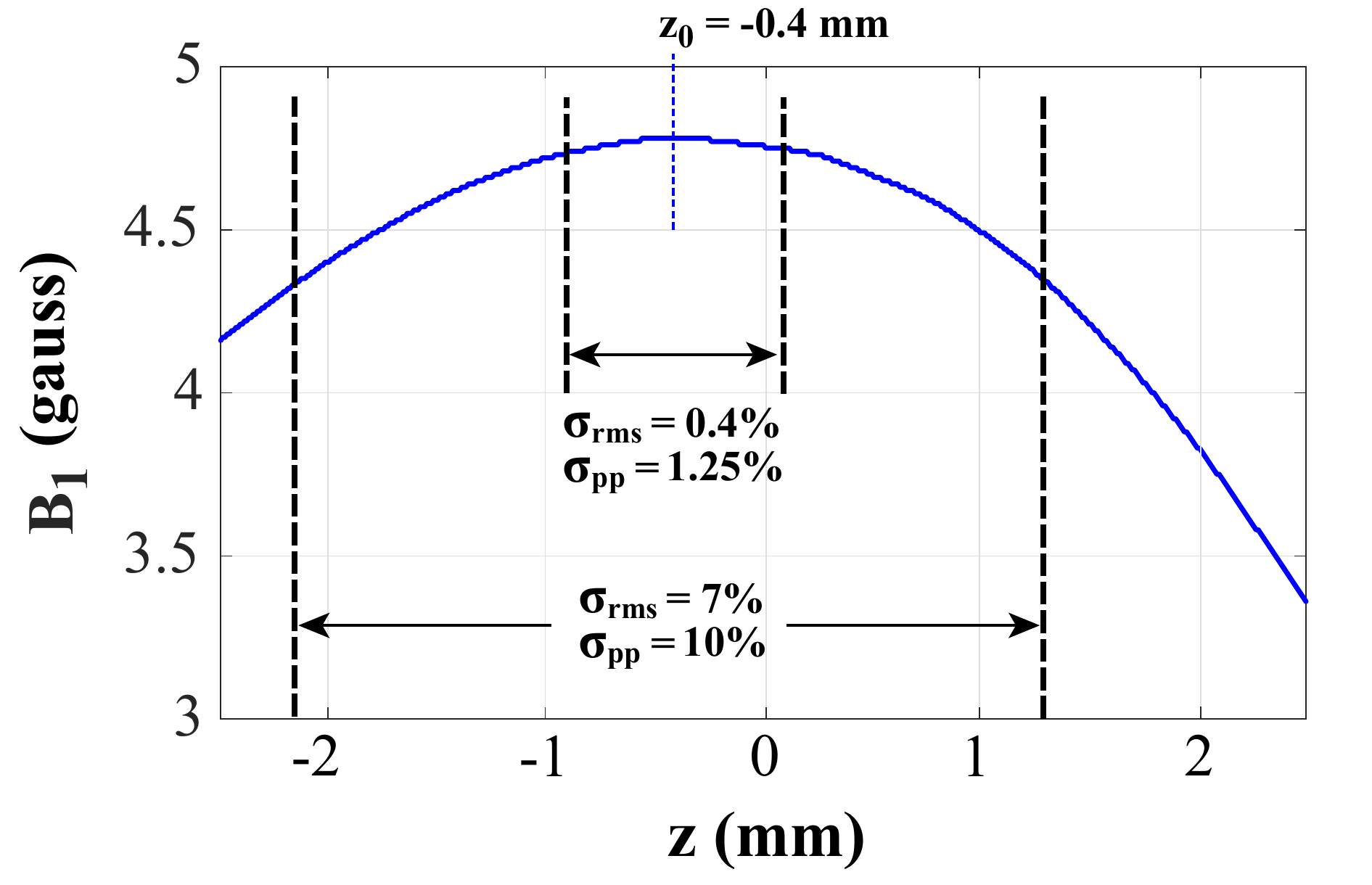}  
\caption{(Color online) \textbf{Simulated $B_1$ field along LGR symmetry axis.} The symmetry plane of the LGR is located at $z = 0$ mm. The edges of the LGR are at $z = \pm 2.5$ mm, and the split-ring resonator is located at position $z = 4.0$ mm. The presence of the split-ring resonator shifts the point of maximal $B_1$ off center to $z_0 = -0.4$ mm.}  \label{ApFig_Z}
\end{center}
\end{figure}

\vspace{-1cm}

\subsection{Smaller Cavity Measurement}\label{smallercavity}

To achieve stronger MW driving, we also designed and fabricated smaller LGR with central loop radius $r_c = 2.5$ mm and $n=4$ outer loops of radius $r_0$ = 2.45 mm, as shown in Fig. \ref{ApFig_two}(a). The naked air-gapped LGR cavity exhibits $f_0 = 4.5$ GHz, similar to the larger LGR design described in the main text. Employing the same exciter antenna from Section \ref{exciterantenna}, we measure $B_1$ = 5.8 G at the center of the smaller LGR device. Figure \ref{ApFig_two}(b) depicts $S_{11}$ for the composite device; the measured 3dB bandwidth $\Delta_\text{3dB}=113$ MHz corresponds to a loaded quality factor $Q_L=25$, and an associated ring-down time of 2.8 ns. 

\begin{figure}[h!]
\centering
\includegraphics[scale = 0.63]{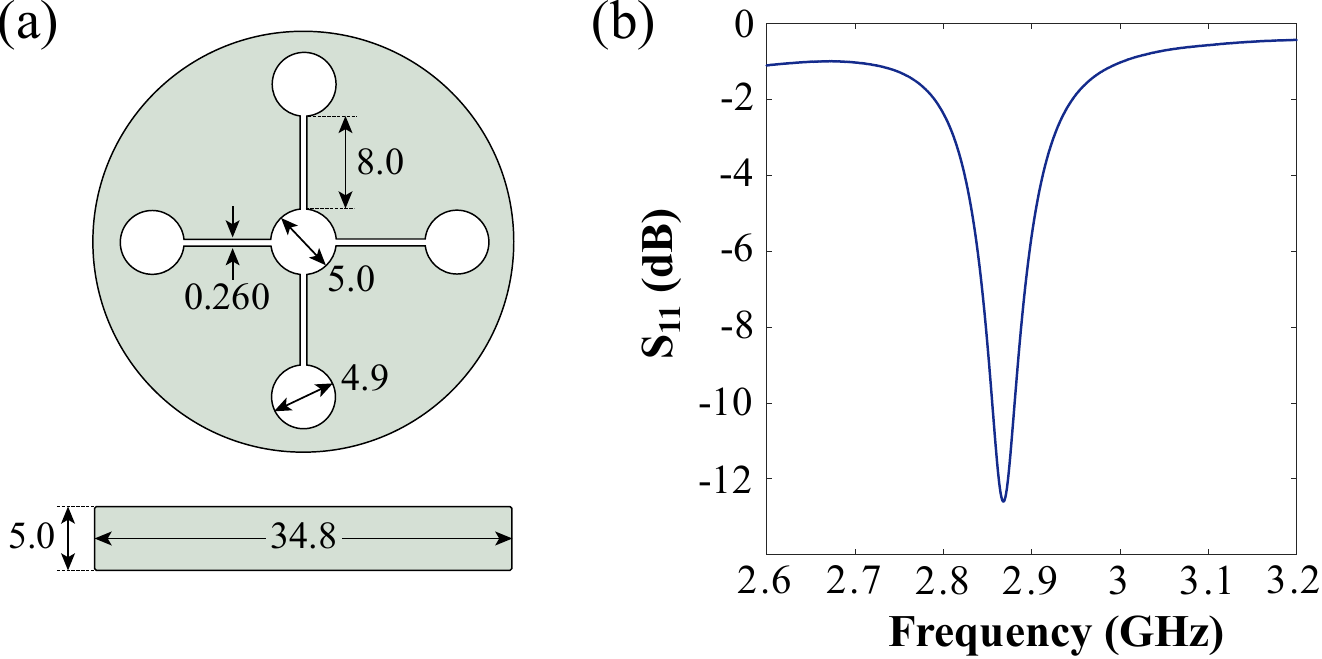}  
\caption{(Color online) \textbf{Smaller LGR design.} \textbf{(a)} Line drawing of smaller LGR with central loop radius $r_c=2.5$ mm as described in Section \ref{smallercavity}. Units are in mm. \textbf{(b)} Measured $S_{11}$ for composite device tuned to $f_0\approx 2.87$ GHz.}   \label{ApFig_two}
\end{figure}

\bibliography{References_Library_Proposal}

\end{document}